\documentclass[twocolumn,prb,showpacs,superscriptaddress,showkeys,preprintnumbers,amsmath,amssymb]{revtex4}

\usepackage{graphicx}
\usepackage{dcolumn}
\usepackage{bm}
\usepackage{multirow}
\usepackage{subscript}
\usepackage{nicefrac}

\begin{document}


\title{Stabilization of the multiferroic spin cycloid in Ni$_3$V$_2$O$_8$ by light Co-doping}

\author{N. Qureshi}

\email[Corresponding author. Electronic
address:~]{qureshi@ph2.uni-koeln.de} \affiliation{$II$.
Physikalisches Institut, Universit\"{a}t zu K\"{o}ln,
Z\"{u}lpicher Strasse 77, D-50937 K\"{o}ln, Germany}

\author{E. Ressouche}

\affiliation{SPSMS, UMR-E CEA/UJF-Grenoble 1, INAC, Grenoble, F-38054, France}

\author{A. A. Mukhin}
\author{V. Yu. Ivanov}

\affiliation{Prokhorov General Physics Institute, Russian Academy
of Sciences, ul. Vavilova 38, Moscow, 119991 Russia}

\author{S. N. Barilo}
\affiliation{Institute of Solid State and Semiconductor Physics, National Academy of Sciences of Belarus, Minsk, 220072 Belarus}

\author{S.V. Shiryaev}
\affiliation{Institute of Solid State and Semiconductor Physics, National Academy of Sciences of Belarus, Minsk, 220072 Belarus}

\author{V. Skumryev}

\affiliation{Instituci\'o Catalana de Recerca i Estudis
Avan\c{c}ats (ICREA), and Departament de F\'isica, Universitat
Aut\'onoma de Barcelona, 08193 Bellaterra, Spain}

\date{\today}

\begin{abstract}

We present macroscopic and neutron diffraction data on multiferroic lightly Co-doped Ni$_3$V$_2$O$_8$. Doping Co into the parent compound suppresses the sequence of four magnetic phase transitions and only two magnetically ordered phases, the paraelectric high temperature incommensurate (HTI) and ferroelectric low temperature incommensurate (LTI), can be observed.  Interestingly, the LTI multiferroic phase with a spiral (cycloidal) magnetic structure is stabilized down to at least 1.8 K, which could be revealed by measurements of the electric polarization and confirmed by neutron diffraction on single crystal samples. The extracted magnetic moments of the LTI phase contain besides the main exchange also fine components of the cycloid allowed by symmetry which result in a small amplitude variation of the magnetic moments along the cycloid propagation due to the site-dependent symmetry properties of the mixed representations. In the HTI phase a finite imaginary part of the spine magnetic moment could be deduced yielding a spin cycloid instead of a purely sinusoidal structure with an opposite spin chirality for different spine spin chains. The magnetic ordering of the cross-tie sites in both phases is different in comparison to the respective ones in the pure Ni compound. A wider temperature stability range of the HTI phase has been observed in comparison to Ni$_3$V$_2$O$_8$ which can be explained by an additional single-ion easy-axis anisotropy due to Co-doping. The larger incommensurability of the Co-doped compounds yields a larger ratio between the competing next-nearest neighbour and nearest neighbour interaction, which is $J_2/J_1$=0.43 (0.47) for a doping level of 7\% (10\%) Co compared to 0.39 in the parent compound.

\end{abstract}


\maketitle

\section{Introduction}
\label{sec:Introduction}

In improper multiferroic materials combining at least two of the ferroic states, e.g. (anti)ferromagnetism and ferroelectricity the ferroelectric polarization is intimately coupled to the magnetic
structure as the first is a direct consequence of the latter in
contrast to proper multiferroics where the origins of both effects
are different. Although the absolute value of the ferroelectric
polarization in improper multiferroics is rather low the
magnetoelectric effect is very high in comparison to the proper
ones. An intensively studied improper multiferroic compound is the kagome staircase system
Ni$_3$V$_2$O$_8$n (NVO) crystallizing in the orthorhombic space group
$Cmce$ (Refs.~\onlinecite{fue1970,sau1973}). In comparison to the
classical kagome structure the symmetry of the kagome staircase is
lowered due to the zig-zag structure and two different
crystallographic sites which are occupied by the magnetic ions
[{\it cross-tie} site (4a), {\it spine} site (8e)].
\begin{figure}
\includegraphics[width=0.49\textwidth]{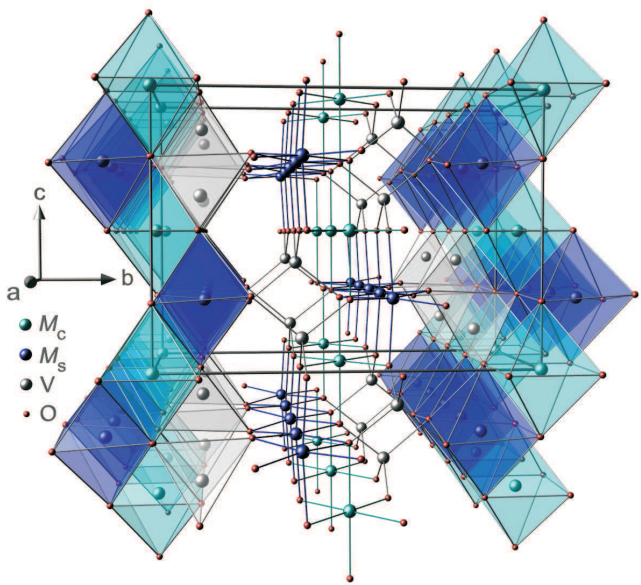}
\caption{\label{fig:structure} (Color online) Crystal structure of
(Co$_x$Ni$_{1-x}$)$_3$V$_2$O$_8$. Polyhedra and bonds have been
drawn to emphasize the structural characteristics. Edge-sharing
$M$O$_6$ ($M$=Co/Ni) octahedra build the kagome staircases which
are separated by VO$_4$ tetrahedra along the {\it b} axis. The
nearest-neighbour magnetic exchange is effectuated by a
90$^{\circ}$ $M$-O-$M$ pathway between a cross-tie (light blue) and a spine ion
or between two spine ions (dark blue).}
\end{figure}
On decreasing temperature the system exhibits a sequence of four
magnetic phase transitions.\cite{rog2002} At 9.1 K
NVO orders antiferromagnetically revealing an
incommensurate sinusoidal spin structure with magnetic moments
predominantly along the {\it a} axis [high-temperature incommensurate (HTI) phase]. Between \mbox{6.3 K} and 3.9 K an
incommensurate spiral magnetic structure in the {\it a-b} plane is
stabilized as a consequence of the geometric frustration [low-temperature incommensurate (LTI) phase]. At
3.9 K the magnetic moments of both sites order in a commensurate canted antiferromagnetic structure,\cite{law2004,law2005,ken2006} before a multi-$\mathbf{k}$ structure is established at 2.3 K where the cross-tie spins exhibit a spiral order.\cite{ehl2013}
The by far most interesting magnetic phase is the spin cycloid as it
breaks inversion symmetry and via the Dzyaloshisnkii-Morya
interaction induces a ferroelectric polarization along the {\it b} axis
therefore rendering the material multiferroic. It has been
demonstrated that the handedness of the spiral can indeed be
switched by reversing an applied electric field.\cite{cab2009}

Substituting Ni with Co gradually lowers the magnetic phase
transition temperature, the variety of magnetic phases and it also changes the
type of magnetic structures as it has been shown by macroscopic
measurements\cite{szy2007,zha2008,zha2011,kum2011} and neutron
diffraction.\cite{qur2006,qur2008/1,qur2008/2}
The application of pressure to NVO reduces the
stability range of the spin spiral until it completely
vanishes.\cite{cha2007} Recently, it has been shown by measurements of the electric polarization\cite{muk2010} that doping a low percentage of Co
into the pure Ni compound stabilizes the multiferroic phase down
to $T$=1.8 K. This behavior is rather surprising as on the one hand the induced chemical
pressure should have the same effect as external pressure and on
the other hand the additional single-ion anisotropy should destroy
the spin spiral as observed for higher Co doping. In this study we address the open question of the magnetic structures of two low Co doped compositions (Ni$_{0.93}$Co$_{0.07}$)$_3$V$_2$O$_8$ and (Ni$_{0.9}$Co$_{0.1}$)$_3$V$_2$O$_8$, which are supposed to be similar to the ones in the parent compound, although no experimental evidence is present so far.

\section{Symmetry analysis}
\label{sec:symmetry}

\begin{figure}
\includegraphics[width=0.4\textwidth]{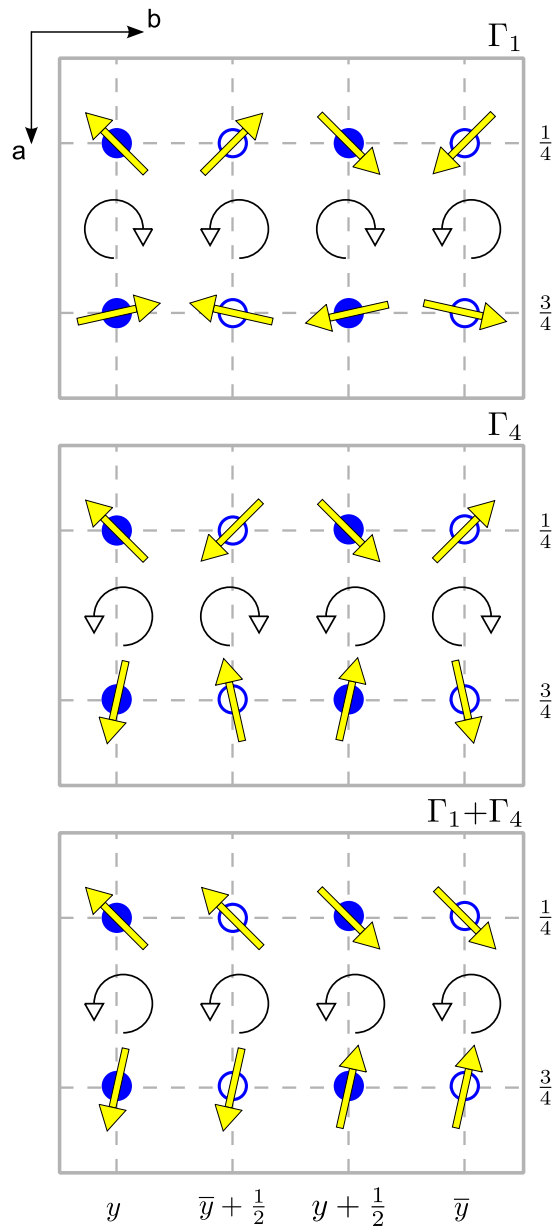}
\caption{\label{fig:spindist} (Color online) Sketch of the crystallographic unit cell (only showing the spine site, filled circles are atoms at $z$=$\frac{1}{4}$, open circles are atoms at $z$=$\frac{3}{4}$) viewed along the $c$ axis for the three spin configurations corresponding to the single representations $\Gamma_1$, $\Gamma_4$ (for both $u_s$=$v_s$) and the mixed representations $\Gamma_1$+$\Gamma_4$ ($u_{4,s}$=$v_{1,s}$ and $u_{1,s}$=$v_{4,s}$=0). Only $\Gamma_1$+$\Gamma_4$ exhibits the same chirality for every spine spin chain along the $a$ axis and therefore induces a ferroelectric polarization along the $b$ axis.  }
\end{figure}

Symmetry adapted
spin configurations have been deduced from representation
analysis. The crystallographic space group connected to a ($q_x$ 0 0)
propagation yields four irreducible representations which have
been tested individually and in combinations on the obtained data.
We denote the irreducible representations from $\Gamma_1$ to
$\Gamma_4$ according to Ref.~\onlinecite{ken2006}. The spin distribution for the different sites ($p$=$c,s$) and their positions $r$ in the magnetic unit cell is determined by the sum of the non-zero basis vectors of the corresponding irreducible representations $\Gamma_n$
\begin{equation}
\label{eq:spindist}
S^{p,r}=\frac{1}{2}\sum\limits_\mathbf{\Gamma_n} \mathbf{\psi}_n\exp{\left[2\pi i \mathbf{q}(\mathbf{R}+\mathbf{r})\right]}+\overline{\mathbf{\psi}}_n\exp{\left[-2\pi i \mathbf{q}(\mathbf{R}+\mathbf{r})\right]}
\end{equation}
where the $\psi_n$ are the complex basis vectors (with $\overline{\mathbf{\psi}}_n$ their complex conjugates) of the $\Gamma_n$ representation for the cross-tie ($p$=$c$, $r$=1-2) and spine spins ($p$=$s$, $r$=1-4) at the respective position $\mathbf{R}$+$\mathbf{r}$ ($\mathbf{R}$ is a lattice vector and $\mathbf{r}$ contains the fractional coordinates) in the crystal (see Tab.~\ref{tab:irrep}). The sum is taken over $\mathbf{q}$ and -$\mathbf{q}$ for which a factor of 1/2 has to be applied. Following Ref.~\onlinecite{ken2006} we use the basis vectors $\psi_n$ in Tab.~\ref{tab:irrep} which take into account additional symmetry operators (in particular, spatial inversion) of the paramagnetic phase that allow the same phase $\phi_{\Gamma_n}$ for all basis vector components of the same $\Gamma_n$ representation. Hence, for a single representation the common phase can be set to zero, while for mixed representations the phase difference between the basis vectors of the respective representations is $\pi/2$. The last column of Tab.~\ref{tab:irrep} shows the resulting basis vectors of such an admixture of representation $\Gamma_1$ ($\psi_1$ shifted by $\pi$/2) to $\Gamma_4$. Although all of the magnetic models presented in Tab.~\ref{tab:irrep} are suited to describe a spine spin cycloid propagating along the $a$ axis, not all of them induce a ferroelectric polarization along the $b$ axis (Note that a sinusoidal magnetic structure does not break inversion symmetry and therefore cannot induce a ferroelectric polarization\cite{che2007}). The polarization direction is determined by the cross product between the spin propagation direction and the axis of spin rotation\cite{che2007} (the latter is given by the cross product between two neighboring spins), which means that in the case of (Ni$_{1-x}$Co$_x$)$_3$V$_2$O$_8$ the spin rotation axis is along $c$. Hence, the magnetic moments of the spine site need to have an $a$ and a $b$ component, from which one of them must be imaginary. Using Eq.~\ref{eq:spindist} with the $C$ translation and the symmetry properties in Tab.~\ref{tab:irrep} one can construct the spine spin distributions for each magnetic model, which is shown in Fig.~\ref{fig:spindist} for a zero $c$ component and a circular envelope in the $a$-$b$ plane.\footnote{Note that this is a simplification of the actual spin cycloid revealed in NVO or in our study. In each magnetic model the real and imaginary parts of the spins in the $a$-$b$ plane have been set equal for all sites resulting in a circular envelope without changes in the spin amplitudes. For the mixed representations $u_{1,s}$ and $v_{4,s}$ have been set to zero which therefore does not represent the small amplitude variation due to different symmetry transformations according to the last column in Tab.~\ref{tab:irrep}. Nevertheless, the simplified models reveal the concept of the induced ferroelectric polarization.} As it can be seen, for the single representations $\Gamma_1$ or $\Gamma_4$ there are two left-handed and two right-handed rotations on the spine spin chains for a propagation along the $a$ axis. According to the definitions above there is an alternation of the spin rotation axis directions which results in an alternation of the polarization vector. In other words, the net electric polarization is zero. In contrast, the sense of spin rotation is the same for all spine chains by using a mixing of representations like $\Gamma_1$+$\Gamma_4$ which results in a non-zero electric polarization as all the oxygen ions (not depicted) are shifted in the same direction due to the inverse Dzyaloshinksii effect. It is a common feature of improper multiferroics, that the multiferroic spin spiral is preceded by a sinusoidal or spiral spin structure whose symmetry does not allow ferroelectricity. MnWO$_4$ (Refs.~\onlinecite{urc2013,sol2013}), AgFeO$_2$ (Ref.~\onlinecite{ter2012}), CuFeO$_2$ (Ref.~\onlinecite{kim2006,ari2007}) and TbMnO$_3$ (Ref.~\onlinecite{ken2005}) among many others undergo a magnetic phase transition at $T_N$ into a spin configuration with magnetic symmetry similar to $\Gamma_1$ or $\Gamma_4$ in Fig.~\ref{fig:spindist}, where in most of the cases a sinusoidally modulated structure is reported, but detailed analysis\cite{urc2013} shows the presence of spin cycloids with opposite chiralities. A few K below a symmetry-breaking magnetic transition takes place into the multiferroic phase in which all the spin cycloids reveal the same chirality. \newline
\begin{table}
\caption{\label{tab:irrep} Basis vectors $\psi_n$ of the irreducible representations $\Gamma_1$, $\Gamma_4$ and $\Gamma_1$+$\Gamma_4$ for the $M^{2+}$ ions [cross-tie (c) and spine (s)] at given fractional coordinates ($x$, $y$, $z$) associated with a propagation vector $\mathbf{q}=(q_x, 0, 0)$. The basis vectors $\psi_n$ result from taking into account additional symmetry operations (in particular, spatial inversion) of the paramagnetic phase and have the same phase factor (see details in Ref.~\onlinecite{ken2006}). For $\Gamma_1$+$\Gamma_4$ a phase factor of $\pi$/2 has been applied to the basis vectors $\psi_1$. The components $u$, $v$ and $w$ connected to the spin $S_{\Gamma_n}^{p,r}$ have been refined according to their constraints (an overline indicates a negative number).}
\begin{ruledtabular}
\begin{tabular}{cccccc}
Site $p$   &   Atom $r$   & ($x$, $y$, $z$) & $\psi_1$ & $\psi_4$ & $\psi_1$+$\psi_4$ \\ \hline \\
$c$ (4a) & 1  & (0, 0, 0)  & $\begin{pmatrix} u_{1,c} \\ 0 \\ 0 \end{pmatrix}$ & $\begin{pmatrix} 0 \\ v_{4,c} \\ w_{4,c} \end{pmatrix} $ & $\begin{pmatrix} iu_{1,c} \\ v_{4,c} \\ w_{4,c} \end{pmatrix} $ \\ \\
         & 2  & (0, $\frac{1}{2}$, $\frac{1}{2}$)  & $\begin{pmatrix} \overline{u}_{1,c} \\ 0 \\ 0 \end{pmatrix}$ & $\begin{pmatrix} 0 \\ \overline{v}_{4,c} \\ w_{4,c} \end{pmatrix} $ & $\begin{pmatrix} i\overline{u}_{1,c} \\ \overline{v}_{4,c} \\ w_{4,c} \end{pmatrix} $   \\ \\
$s$ (8e) & 1  & ($\frac{1}{4}$, $y$, $\frac{1}{4}$)  & $\begin{pmatrix} iu_{1,s} \\ v_{1,s} \\ iw_{1,s} \end{pmatrix}$ & $\begin{pmatrix} u_{4,s} \\ iv_{4,s} \\w_{4,s} \end{pmatrix}$  & $\begin{pmatrix} \overline{u}_{1,s}+u_{4,s} \\ i{v}_{1,s}+iv_{4,s} \\ \overline{w}_{1,s}+w_{4,s} \end{pmatrix} $  \\ \\
& 2  & ($\frac{1}{4}$, $\overline{y}$, $\frac{3}{4}$)  & $\begin{pmatrix} iu_{1,s} \\ \overline{v}_{1,s} \\ i\overline{w}_{1,s} \end{pmatrix}$ & $\begin{pmatrix} \overline{u}_{4,s} \\ iv_{4,s} \\ w_{4,s} \end{pmatrix} $ & $\begin{pmatrix} \overline{u}_{1,s}+\overline{u}_{4,s} \\ i\overline{v}_{1,s}+iv_{4,s} \\ w_{1,s}+w_{4,s} \end{pmatrix} $ \\ \\
         & 3  & ($\frac{1}{4}$, $\overline{y}$+$\frac{1}{2}$, $\frac{3}{4}$)  & $\begin{pmatrix} i\overline{u}_{1,s} \\ v_{1,s} \\ i\overline{w}_{1,s} \end{pmatrix}$ & $\begin{pmatrix} u_{4,s} \\ i\overline{v}_{4,s} \\ w_{4,s} \end{pmatrix} $ & $\begin{pmatrix} {u}_{1,s}+{u}_{4,s} \\ i{v}_{1,s}+i\overline{v}_{4,s} \\ w_{1,s}+w_{4,s} \end{pmatrix} $   \\ \\
& 4  & ($\frac{1}{4}$, $y$+$\frac{1}{2}$, $\frac{1}{4}$)   & $\begin{pmatrix} i\overline{u}_{1,s} \\ \overline{v}_{1,s} \\ iw_{1,s} \end{pmatrix}$ & $\begin{pmatrix} \overline{u}_{4,s} \\ i\overline{v}_{4,s} \\ w_{4,s} \end{pmatrix} $ & $\begin{pmatrix} {u}_{1,s}+\overline{u}_{4,s} \\ i\overline{v}_{1,s}+i\overline{v}_{4,s} \\ \overline{w}_{1,s}+w_{4,s} \end{pmatrix} $
\end{tabular}
\end{ruledtabular}
\end{table}

\section{Experimental}
\label{sec:Experimental}

Samples with dimensions suitable for the particular measurements and with different crystallographic orientations were cut from the
same crystals used in Ref.~\onlinecite{muk2010}.
The magnetization was measured as a function of temperature
using a superconducting quantum interference device
(SQUID) magnetometer from Quantum Design. A magnetic
field of 1 kOe was applied within a few degrees of accuracy
along  the
principle crystallographic directions of the crystals (shaped as
a 3x2x5 mm$^3$ parallelepiped along the crystallographic $a$, $b$ and $c$ axes, respectively). The data presented here
are not corrected for the demagnetizing field effect. The electric polarization was studied along the $b$ axis by pyroelectric measurements using a Keithley 6517A electrometer.
The neutron diffraction experiment has been carried out at the
four-circle diffractometer D10 (ILL, Grenoble) using a wavelength
of 1.26~\AA{} from a Cu(200) monochromator. The same crystals were used as for the bulk magnetic measurements. The nuclear structures
of both compounds have been investigated within the paramagnetic
phase at 15 K, while the magnetic structures have been studied at
2 K and \mbox{7 K} according to the magnetic phase transitions seen by
the macroscopic methods.

\section{Results and discussion}
\label{sec:results}

\subsection{Macroscopic measurements}
\label{sec:macroscopic}

\begin{figure}
\includegraphics[width=0.49\textwidth]{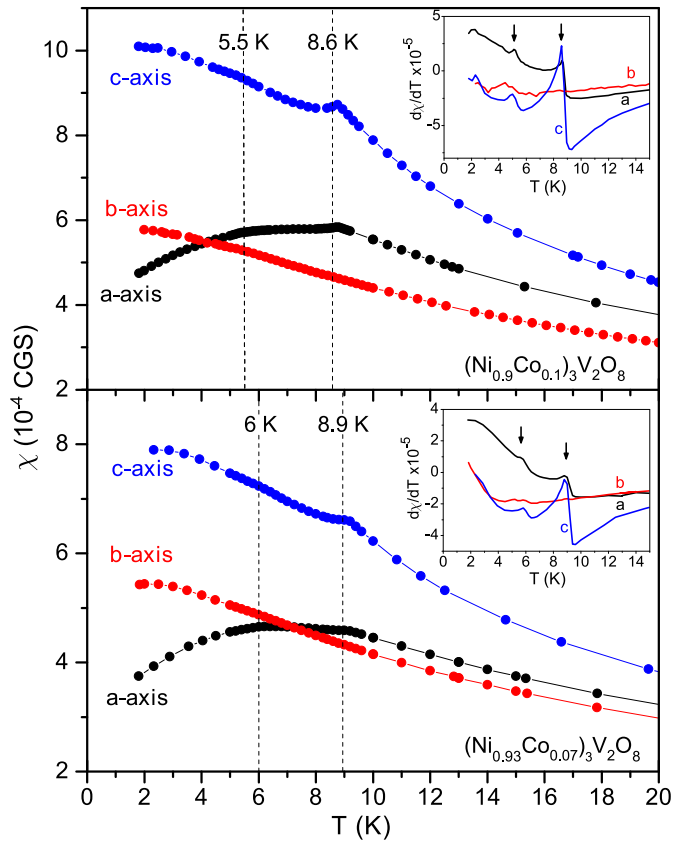}
\caption{\label{fig:susc} (Color online) Temperature dependence of the magnetic susceptibility of (Ni$_{0.9}$Co$_{0.1}$)$_3$V$_2$O$_8$ (upper pannel) and (Ni$_{0.93}$Co$_{0.07}$)$_3$V$_2$O$_8$ (lower pannel) for an external field of 1 kOe along the principal crystallographic directions. The derivatives of the susceptibility are plotted in the insets. The arrows and the dashed lines indicate susceptibility changes attributed to magnetic phase transitions. }
\end{figure}

\begin{figure}
\includegraphics[width=0.49\textwidth]{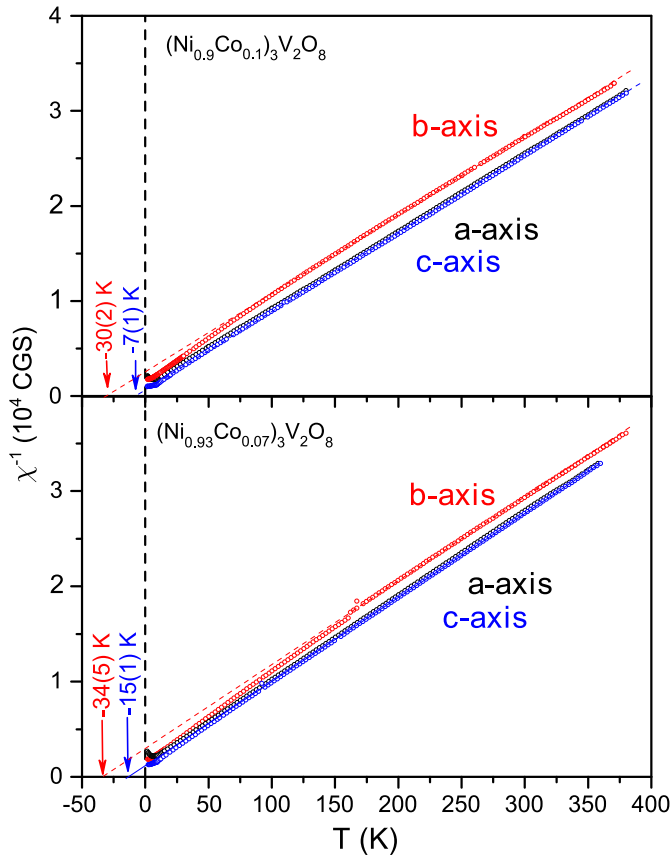}
\caption{\label{fig:invsusc} (Color online) Temperature dependence of the inverse magnetic susceptibility of (Ni$_{0.9}$Co$_{0.1}$)$_3$V$_2$O$_8$ (upper pannel) and (Ni$_{0.93}$Co$_{0.07}$)$_3$V$_2$O$_8$ (lower pannel) for an external field of 1 kOe along the principal crystallographic directions. The arrows indicate the asymptotic paramagnetic temperatures $\theta_P$.}
\end{figure}

\begin{figure}
\includegraphics[width=0.49\textwidth]{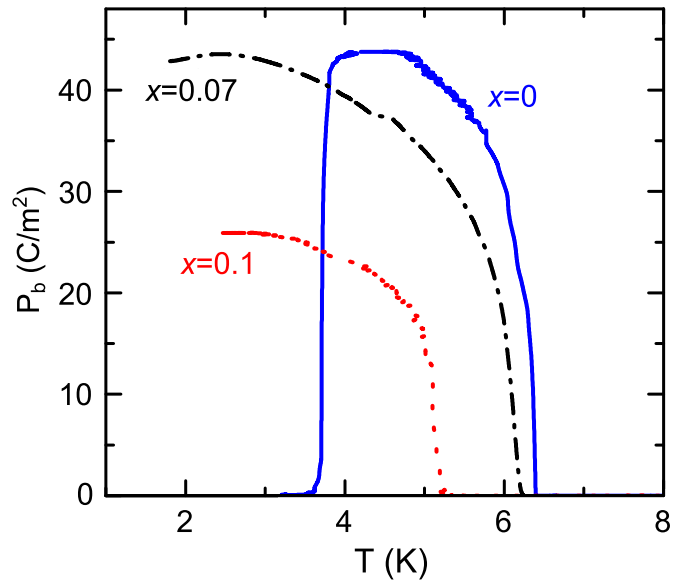}
\caption{\label{fig:epol} (Color online) Temperature dependence of the electric polarization $P_b(T)$ in pure and Co-doped nickel vanadates, illustrating a suppression of the low temperature weak ferromagnetic phase for the doped compounds in comparison to NVO ($x$=0).}
\end{figure}

The temperature dependence of the magnetic susceptibility along the principal crystallographic directions at low temperatures is shown in Fig.~\ref{fig:susc}. For both the $x=0.07$ and $x=0.1$ compositions, a peculiarity is clearly visible in the $a$ and $c$ axis susceptibilities at \mbox{9.1(1) K} and 8.7(1) K, respectively, but not on the $b$ axis susceptibility, very similar to the behavior in the Co-richer compound (Ni$_{0.5}$Co$_{0.5}$)$_3$V$_2$O$_8$ (Ref.~\onlinecite{qur2008/2}). This is better manifested in the insets of Fig.~\ref{fig:susc}, where the derivatives of the susceptibility are plotted. Another clear peculiarity is observed at lower temperatures for both compositions, at 6.0(2) K and 5.5(2) K, respectively, in magnetic fields along the $a$ and the $c$ axis. While the high temperature peculiarity could be naturally attributed to the N\'eel temperature, which is gradually decreased upon Co doping\cite{qur2006,qur2008/1} ($T_N$=9.1 K for the parent compound), the low temperature susceptibility anomaly signals a change in the magnetic structure.\newline
As seen from Fig.~\ref{fig:invsusc}, for both compositions the anisotropy observed at and below $T_N$ persists in the paramagnetic region with all the three inverse susceptibilities becoming parallel straight lines at high enough temperatures where the Curie-Weiss law applies. However, while at high temperatures the $a$ and $c$ axis susceptibilities become almost the same, they remain different from the one along the $b$ axis. Note that, in contrast, the paramagnetic susceptibility of the undoped compound is rather isotropic.\cite{szy2009}
The observed anisotropy reflects the significant difference in the asymptotic paramagnetic temperatures $\theta_P$ obtained for the three directions: \mbox{$\theta_P$=-34(5) K} along the $b$ axis and \mbox{$\theta_P$=-15(1) K} along the $a$ and $c$ axes of (Ni$_{0.93}$Co$_{0.07}$)$_3$V$_2$O$_8$; \mbox{$\theta_P$=-30(2) K} along the $b$ axis and \mbox{$\theta_P$=-7(1) K} along the $a$ and $c$ axes of (Ni$_{0.9}$Co$_{0.1}$)$_3$V$_2$O$_8$. A tendency for a weakening of the exchange interactions with an increase of Co doping is seen. Within the experimental error, the effective magnetic moment was found to be the same for all the orientations: 6.1(1)~$\mu$\textsubscript{B}/f.u. in (Ni$_{0.93}$Co$_{0.07}$)$_3$V$_2$O$_8$ and 6.3(1)~$\mu$\textsubscript{B}/f.u. (Ni$_{0.9}$Co$_{0.1}$)$_3$V$_2$O$_8$ (note that vanadium is in V$^{5+}$ valence and diamagnetic). Those values are bigger than the ones based on the spin only contribution from the $3d$ ions and suggest that the orbital moments are not completely frozen. Figure 4 shows the temperature dependence of the electric polarization $P(T)$ of pure and doped nickel vanadates measured after the preliminary cooling of the samples in an electric field of the order of 1 kV/cm. According to the figure, the behavior of the polarization in pure nickel vanadate is in qualitative agreement with the data reported in Ref.~\onlinecite{zha2011}. The presence of an electrical polarization has been shown\cite{law2004,law2005,ken2006} to be caused by the inverse Dzyaloshinksii effect due to a spiral magnetic structure which breaks inversion symmetry. Therefore, the electric polarization serves as an unambiguous proof for the transitions from the HTI to the multiferroic LTI phase and from the LTI to the commensurate phases at lower temperature. The experimental data in Fig.~\ref{fig:epol} shows that for the doped compounds, the transition temperature from the HTI to the LTI phase is reduced as compared with the pure one (down to 6.1 and 5.3 K for $x$=0.1 and 0.07, respectively) which is in a reasonable agreement with the magnetic data. As it is seen from the figure, in contrast to NVO, the polarization in lightly doped vanadates does not disappear as the temperature is reduced, which discards the presence of an additional magnetic phase above 1.8 K.

\subsection{Nuclear structure}
\label{nuclear}

The nuclear structure refinement was based on 800 measured unique
reflections of the crystallographic space group $Cmce$. The
integrated intensities were corrected for absorption applying the
transmission factor integral $\exp[-\mu (\tau_{in}+\tau_{out})]$
by using subroutines of the \textsc{Cambridge Crystallographic
Subroutine Library}\cite{ccsl} [$\tau_{in}$ and $\tau_{out}$
represent the path lengths of the beam inside the crystal before
and after the diffraction process, $\mu$ is the linear absorption
coefficient, which is 0.129 cm$^{-1}$ and 0.145 cm$^{-1}$ for
(Ni$_{0.93}$Co$_{0.07}$)$_3$V$_2$O$_8$ and
(Ni$_{0.9}$Co$_{0.1}$)$_3$V$_2$O$_8$, respectively]. The
fractional coordinates and isotropic temperature factors have been
refined for each atom except for V whose position and Debye-Waller factor have been fixed
to the one observed by X-ray measurements due to its transparency
for neutrons. Further refined parameters were the Co/Ni occupation
on the $4a$ and $8e$ site and the extinction parameters according
to an empiric ShelX-like model.\cite{lar1970} All refined
structural parameters are shown in Tab.~\ref{tab:structure}.

\begin{table}
\caption{\label{tab:structure} Refined nuclear structure
parameters within the $Cmce$ space group for
(Ni$_{1-x}$Co$_x$)$_3$V$_2$O$_8$ with a nominal Co-doping of $x_{nom}=0.07$ and
$x_{nom}=0.10$. $x_{ref}$ specifies the refined amount of doping. The extinction parameters $x_{ij}$ are the diagonal entries of a tensor used to calculate the extinction factor.}
\begin{ruledtabular}
\begin{tabular}{cccc}
          &   & $x_{nom} = 0.07$ & $x_{nom} = 0.10$ \\ \hline
cell & $a$ (\AA)  & 5.9330(5)  & 5.9362(4)    \\
     & $b$ (\AA)  & 11.3883(9)  & 11.3896(6)     \\
     & $c$ (\AA)  & 8.2315(8) & 8.2289(6)     \\
Co1/Ni1 & occ. (\%)        & 13(2)/87(2)  & 18(2)/82(2) \\
        & $B$ (\AA$^2$)    & 0.04(12)  & 0.0(1)   \\
Co2/Ni2 & $y$   &  0.1303(1)          &   0.1302(1)         \\
        & occ. (\%)       & 6(1)/94(1)  & 13(2)/87(2) \\
        & $B$ (\AA$^2$)    & 0.10(6)  & 0.1(1)   \\
$x_{ref}$  & &  0.08(2)       &   0.15(2)         \\
O1 & $y$   &  0.2490(2) &  0.2490(3) \\
   & $z$   &  0.2685(3) &  0.2682(5)  \\
        & $B$ (\AA$^2$)    & 0.16(7)  & 0.2(1)   \\
O2 & $y$   &  0.0011(2)  & 0.0011(3)   \\
   & $z$   &  0.2448(4)  & 0.2445(6)  \\
        & $B$ (\AA$^2$)    & 0.11(7)  & 0.2(1)   \\
O3 & $x$   &  0.2650(5)  &  0.2655(4)  \\
   & $y$   &  0.1191(2) & 0.1190(2)  \\
   & $z$   &  0.0000(3) & 0.9996(4)  \\
        & $B$ (\AA$^2$)    & 0.22(7)  & 0.3(1)   \\
ext.   & $x_{11}$            & 0.07(2)  & 0.10(1) \\
       & $x_{22}$            & 0.0107(8)  & 0.039(4)  \\
       & $x_{33}$            & 0.062(3)  & 0.11(2)\\
       \hline
       & $R_F$ (\%)    & 4.10 & 4.98\\
       & $\chi^2$      & 17.2 & 17.0 \\
\end{tabular}
\end{ruledtabular}
\end{table}

The structural parameters show to be as expected including a
slightly higher Co concentration with the preference to occupy the
more symmetric $4a$ site like it has already been
observed for the mixed series (Co$_x$Ni$_{1-x}$)$_3$V$_2$O$_8$.\cite{qur2008/1,qur2008/2} \newline 

\subsection{Magnetic structures}

For each magnetically
ordered phase a number of 200 unique reflections has been
recorded, integrated and corrected for absorption. Both compounds
reveal two magnetically ordered phases, which prove to be
incommensurately modulated by a propagation of type ($q_x$ 0 0)
in analogy to the pure Ni compound ($q_x$ is composition dependent). 

\subsubsection{HTI phase}

\begin{figure*}
\includegraphics[width=0.49\textwidth]{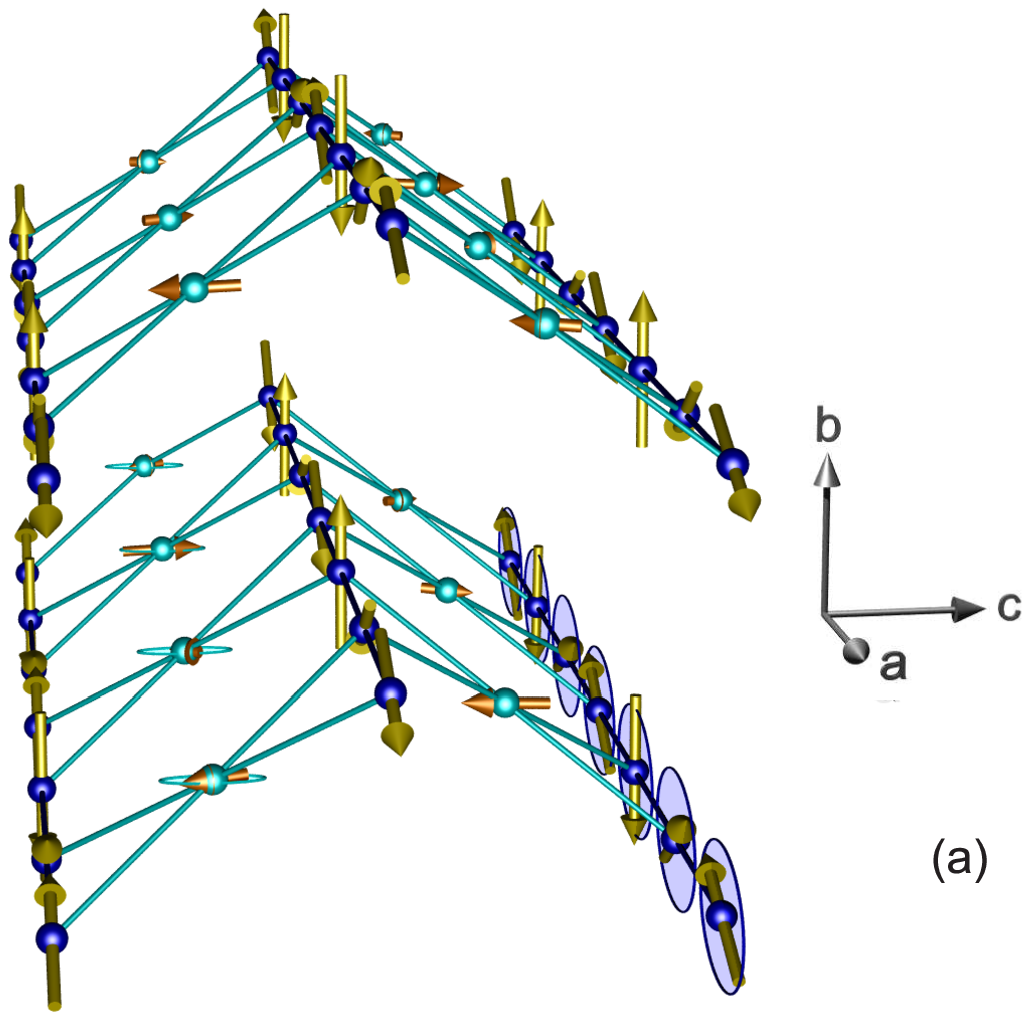}
\includegraphics[width=0.49\textwidth]{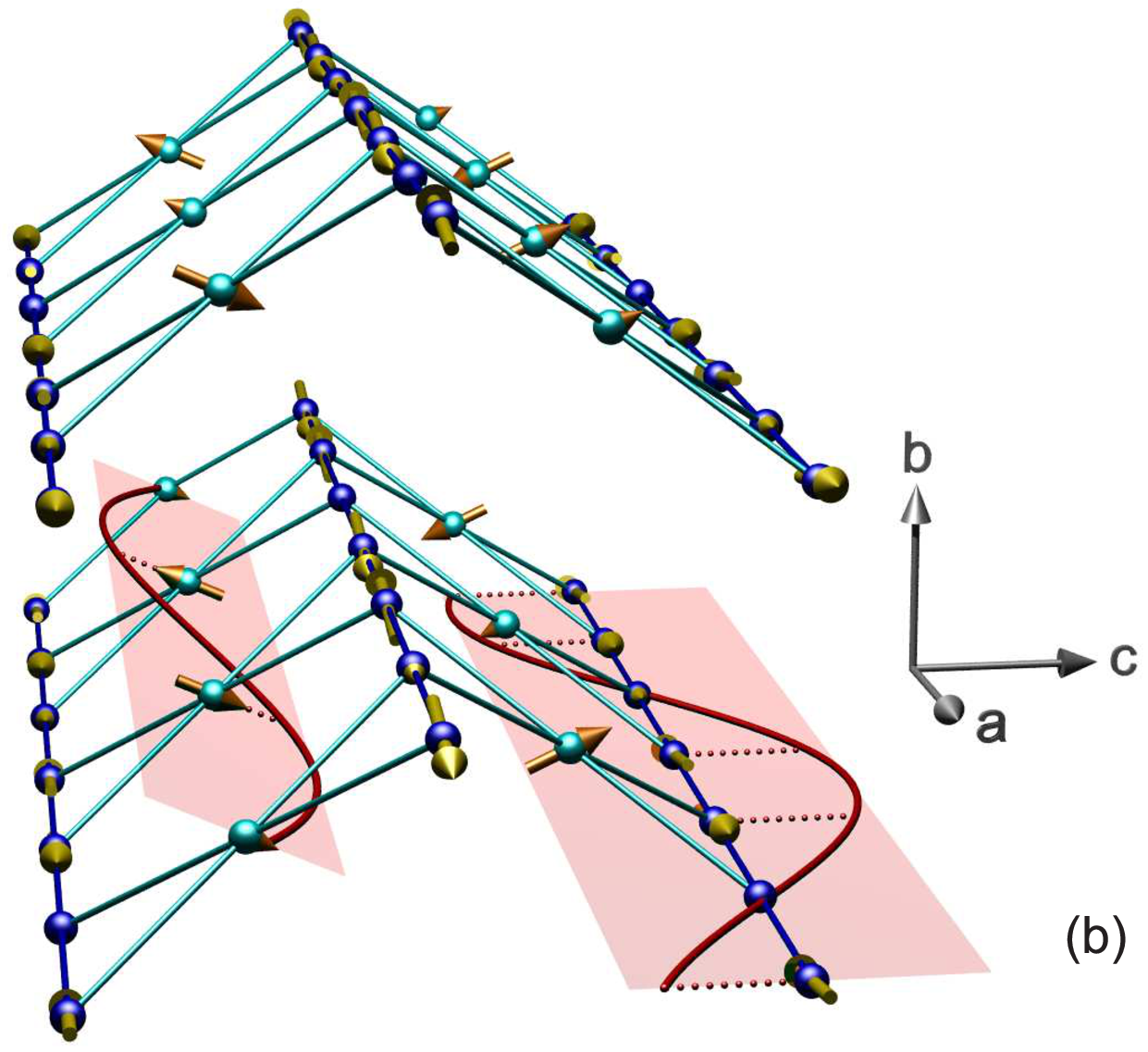}
\caption{\label{fig:magstruct} (Color online) Magnetic structures of (Ni$_{1-x}$Co$_{x}$)$_3$V$_2$O$_8$ for $x$=0.07 and 0.10. (a) At $T$=7 K the cross-tie moments are sinusoidally modulated, while the spine moments describe a rotation around the $c$ axis with a strongly elongated elliptical envelope, which is close to a sinusoidal modulation (the size of a magnetic moment at a particular position is emphasized by a [red] sine curve with a wavelength of $a$/$q_x$). The spine spins predominantly point along the $a$ axis with a small imaginary $b$ component and an essentially zero $c$ component, whereas the cross-tie spins lie within the {\it b-c} plane. (b) at $T$=2 K a spin cycloid is established where the magnetic moments of the spine site rotate within the {\it a-b} plane [emphasized by the (blue) envelope] propagating along the propagation vector ($q_x$ 0 0).  Two neighboring kagome staircases are depicted showing the antiparallel alignment of spine spins of adjacent layers. The rotation of the cross-tie spins traces a narrow ellipse within the {\it a-c} plane. }
\end{figure*}

The observed magnetic scattering intensities at \mbox{7 K} could well be described by the single
irreducible representation $\Gamma_4$ ($\phi_{\Gamma_4}$=0) for both investigated compounds. The resulting components are shown in Tab.~\ref{tab:magresults}. The cross-tie moments are not completely frustrated like in the undoped compound, so that their non-zero real components result in a sinusoidal modulation along the propagation vector $\mathbf{q}$=(0.32 0 0) [$\mathbf{q}$=(0.30 0 0)] for $x$=0.10 ($x$=0.07) with the spins lying within the {\it b-c} plane. Collinear chains of cross-tie spins are formed along the $a$ axis, where the $b$ component of the spin alternates from chain to chain along the [001] direction.  The imaginary $b$ component of the spine spins may induce a spin cycloid, however, it is small compared to the strong real component along the $a$ axis, while the $c$ component is essentially zero. This results in a spin cycloid with a strongly elongated elliptical envelope which is therefore close to a sinusoidal modulation. The configuration of the spine spins can therefore be described as close to collinear along the $a$ axis. The magnetic structure is visualized in Fig.~\ref{fig:magstruct}(a) showing two neighboring kagome staircases. Note that the spin configuration according to the single irreducible representation $\Gamma_4$ does not induce a ferroelectric polarization as it has been described in Sec.~\ref{sec:symmetry} and shown in Fig.~\ref{fig:spindist}.
\begin{table*}
\caption{\label{tab:magresults} Refined components of the basis vectors for spine and cross-tie positions $r$=1 given in Bohr magnetons for the two magnetic structure models describing the data at $T$=2 K ($\Gamma_1+\Gamma_4$) and $T$=7 K ($\Gamma_4$ only), respectively. Note that the basis functions of $\Gamma_1$ are shifted by the phase factor $\pi/2$ with respect to those shown in Tab.~\ref{tab:irrep}. }
\begin{ruledtabular}
\begin{tabular}{cccc}
$T(K)$ & $S_{\Gamma_n}^{p,r}$  &   $x$=0.07  & $x$=0.10  \\ \hline \\
& $S_{\Gamma_1}^{c,1}$  &   [-0.13(9)$i$, 0, 0] & [-0.08(7)$i$, 0, 0]  \\ \\
 & $S_{\Gamma_4}^{c,1}$    & [0, 0.07(7), 0.12(5)] &  [0, -0.05(6), 0.18(4)]\\ \\
2 &  $S_{\Gamma_1}^{s,1}$   & [ -0.04(6), 1.58(3)$i$, -0.02(6)] &  [-0.07(5), 1.39(3)$i$, -0.02(5)]  \\ \\
&  $S_{\Gamma_4}^{s,1}$   & [2.08(3),  0.141(2)$i$, 0.013(2)] &  [1.99(2), 0.122(1)$i$, -0.013(1)]  \\ \\
&      $\chi^2$ ($R_F$)        & 9.30 (7.10)  &  4.36 (6.61)  \\\hline \\
&      $S_{\Gamma_4}^{c,1}$   & [0, -0.10(3), 0.07(2)]  & [0, -0.12(3), 0.17(3)]  \\ \\
7&  $S_{\Gamma_4}^{s,1}$   & [1.63(2), 0.11(2)$i$, 0.01(5)]  & [1.49(1), 0.09(2)$i$, -0.01(5)] \\ \\
&        $\chi^2$ ($R_F$)        & 4.93 (11.6)  & 2.49 (9.33)  \\

\end{tabular}
\end{ruledtabular}
\end{table*}

\subsubsection{$LTI$ phase}
As known from the pure Ni compound a mixing of the two representations $\Gamma_1$ and $\Gamma_4$ is needed to describe the LTI phase, where the basis functions $\psi_1$ are shifted by $\pi/2$ with respect to $\psi_4$. In a first step the coefficients of the basis vectors $\psi_n$ have been refined, while restricting the real and the imaginary part of the magnetic moments of both sites to be perpendicular, of equal size and within the {\it a-b} plane. This corresponds to a spin cycloid perpendicular to the $c$ axis with a circular envelope. The agreement of the imposed model with the data was quite good, but could substantially be improved by releasing the constraint on the size of the real and imaginary parts resulting in an elliptical envelope just as reported in Ref.~\onlinecite{ken2006}.
Furthermore, one may suggest that for the spine spins the ratio between the
main (exchange) component $u_{4,s}$ of the basis vector and the weak (relativistic) ones $iv_{4,s}$ and $w_{4,s}$ in the HTI phase remains the same in the LTI phase (i.e. $\frac{u_{4,s,HTI}}{v_{4,s,HTI}}$=$\frac{u_{4,s,LTI}}{v_{4,s,LTI}}$ and $\frac{u_{4,s,HTI}}{w_{4,s,HTI}}$=$\frac{u_{4,s,LTI}}{w_{4,s,LTI}}$) due to the presence of the mixed terms $u_{4,s}(v_{4,s})^*$ and  $u_{4,s}(w_{4,s})^*$ allowed by symmetry in the Landau thermodynamic potential which results in the proportionalities $v_{4,s}\sim u_{4,s}$ and $w_{4,s}\sim u_{4,s}$.
As this could be confirmed in the first refinement steps - although with limited precision - this ratio has been constrained. A further reduction of $\chi^2$ was achieved by introducing a $c$-component on the real part of $S^c_{\Gamma_4}$. It turned out that for the $x$=0.10 compound the only significant component of the cross-tie moment points along the $c$ axis essentially resulting in a sinusoidal modulation of collinear spins. The spine spins instead form a spin cycloid propagating along the $a$ axis with its spin rotation axis along $c$. Due to the same chirality of each spine spin chain (see Fig.~\ref{fig:spindist}) a uniform electric polarization is induced along $b$ via the Dzyaloshinskii-Moriya interaction (see Sec.~\ref{sec:symmetry}). Compared to NVO the spine spin configuration is qualitatively the same and only the values of the magnetic moments differ slightly. Concerning the cross-tie site, there is a change from a spin cycloid within the $a$-$b$ plane in the undoped compound to a presumably collinear sinusoidally modulated structure with $\mu\parallel c$ for $x=0.10$ as the $a$ and $b$ components do not reveal significant non-zero values. For $x=0.07$ a finite $a$-component of -0.13(9)~$\mu$\textsubscript{B} has been refined, for which a spin spiral within the $a-c$ plane may be possible although the data does not permit a clear statement. The magnetic structure of the LTI phase is visualized in Fig.~\ref{fig:magstruct}(b). 

It can be seen in Tab.~\ref{tab:magresults} that the major spine spin components are the imaginary $b$ component of $\Gamma_1$ and the real $a$ component of $\Gamma_4$. However, the real $a$ component of $\Gamma_1$ and the imaginary $b$ component of $\Gamma_4$ show small but reliable values in comparison to Ref.~\onlinecite{ken2006} (excepting $u_{1,s}$ for $x$=0.07). 
The existence of those fine components of the spine spins in this phase makes the cycloidal structure locally different resulting in a non-equivalence of the spin amplitudes. This is due to the fact the two irreducible representations transform the spin properties according to different symmetry operators, for which the sum or the difference of two spin components can be realized depending on the site (see Tab.~\ref{tab:irrep}). This situation is very similar to the recent finding of non-equivalent spin cycloids in various Mn$^{2+}$ sites in the ferrorelectric phase of MnWO$_4$ multiferroics.\cite{urc2013}


\section{Conclusion}
\label{sec:discussion}

We have carried out magnetization and neutron diffraction experiments documenting the magnetically ordered phases and the respective transition temperatures of two multiferroic compounds (Ni$_{0.93}$Co$_{0.07}$)$_3$V$_2$O$_8$ and (Ni$_{0.9}$Co$_{0.1}$)$_3$V$_2$O$_8$. Substituting more than 3.5\% Ni by Co suppresses the sequence of four magnetic phase transitions, reduces the N\'eel temperature and makes the propagation vector temperature independent.\cite{qur2008/1} Our macroscopic measurements show clearly the existence of two magnetic phase transitions above 1.8 K. The corresponding phases are antiferromagnetic, from which the low-temperature phase exhibits a non-zero electric polarization as shown in Ref.~\onlinecite{muk2010}.
The magnetic structure determination by means of neutron diffraction revealed that the two magnetically ordered phases correspond to the HTI and LTI phases of the parent compound.\cite{law2004} Although the spin configurations correspond to the same irreducible representations as the respective NVO structures, small differences are observable concerning the cross-tie spin order due to different spin components which, however, leave the symmetry properties invariant. In the HTI phase the cross-tie spins are not completely frustrated as in the pure Ni compound exhibiting a magnetic moment of 0.21(3)~$\mu$\textsubscript{B} for the $x$=0.1 compound (0.13(3)~$\mu$\textsubscript{B} for $x$=0.07), which is oriented within the {\it b-c} plane. Due to the finite imaginary part of the magnetic moment the spine spins form a cycloid instead of a purely sinusoidal structure with an opposite spin chirality for different spine chains for which this magnetic structure does not induce a ferroelectric polarization. Such a fine structure, which is determined by the weak anisotropic (relativistic) magnetic interactions and distorts the main magnetic structure on the different magnetic sites, has also been observed in the LTI phase. These fine components result in a non-equivalence of the spin amplitudes along the cycloid propagation direction. A similar fine structure has been observed recently in the multiferroic compound MnWO$_4$,\cite{urc2013} where also the symmetry of the magnetic structures reveal exactly the same pattern as presented here, i.e. spin spirals with opposite chiralities in the close to sinusoidal phase and spin spirals with the same chirality in the multiferroic phase.\cite{urc2013,sol2013} Our results reveal that in the LTI phase the cross-tie spins do not exhibit the same cycloid as the spine spins. Instead, a close to sinusoidal modulation is present in the $x=0.10$ compound, where the only significant spin component points along the $c$ axis, very similar to the HTI phase. For $x=0.07$ a finite $a$ component has been refined besides the significant $c$ component, for which also a spin spiral within the $a$-$c$ plane may be possible. The average magnetic moments and their orientation in the LTI and HTI phases are quite comparable in the doped compounds and only the type of the propagation changes from a sinusoidal modulation to a spin cycloid. Therefore, the LTI and HTI phases resemble each other much more than in NVO, where especially the cross-tie moment changes from almost fully frustrated to roughly 1.4~$\mu$\textsubscript{B} (Ref.~\onlinecite{ken2006}). A reason for the reduced cross-tie magnetic moments in the doped compounds could be the preferential occupation of Co on this site which seems to affect the ordered moment more than it does on the spine site. On a microscopic level the statistical occupation of the Co$^{2+}$ ions would lead to randomly oriented isotropic exchange fields at the cross-tie position created by the four surrounding spine-sites depending on the nearest neighbour atom type. These fields should locally magnetize the cross-tie spins and suppress the regular magnetic modes corresponding to the $\Gamma_1$ and $\Gamma_4$ representations. \newline
It is a surprising fact that Co-doping, applying a certain amount of chemical pressure, stabilizes the spin cycloid down to low temperature whereas external pressure reduces its stability range.\cite{cha2007} Furthermore, the additional single-ion anisotropy due to the Co$^{2+}$ ions should favor a collinear antiferromagnetic ordering along the $a$-axis similar to the parent compound Co$_3$V$_2$O$_8$.\cite{che2006} From this point of view the observed existence of the cycloidal LTI phase down to low temperatures seems unexpected. We suggest that it could be explained by a change of the fine balance between the exchange interactions and the single-ion anisotropy in the Co-doped compounds. In pure NVO the exchange interactions between two Ni spins along the spine chains are responsible for the cycloidal LTI and sinusoidal HTI phases where the competing exchange between the nearest ($J_1
>0$) and next-nearest ($J_2>0$) neighbors is important.\cite{ken2006} In particular, these interactions determine the incommensurate wave vector of the magnetic structures according to $\cos\left[(1-q_x)\pi\right]\approx - J_1/4J_2$. The low temperature transition into the collinear commensurate antiferromagnetic phase ($C$ phase) is determined by a competition between the single-ion easy-axis anisotropy $K$ stabilizing the $a$-axis spin direction and the exchange interactions which occurs in a narrow range of the $K$, $J_1$ and $J_2$ values according to the simulation given in Ref.~\onlinecite{ken2006}. An increase of the effective anisotropy $K$ due to Co doping has to be compensated by an increasing $J_2/J_1$ ratio in order to suppress the low temperature transition into the $C$ phase (see Fig.~20 in Ref.~\onlinecite{ken2006}), while $K/J_1$ stays inferior to a critical value of 0.5 (see Fig.~19 in Ref.~\onlinecite{ken2006}). Indeed, our experimental observations confirm these assumptions. As mentioned above, the incommensurability is a result of the competition between $J_1$ and $J_2$. Using $q_x=0.28$ of the parent compound yields a $J_2/J_1$ ratio of 0.39 while for the Co doped compounds we can estimate from our data an increase of this ratio up to 0.43 for $x$=0.07 and 0.47 for $x$=0.10, respectively, which confirms the stabilization of the LTI phase. At the same time we suggest that $J_1$ increases in order to compensate the increasing $K$ according to $K/J_1\leq 0.5$. As stated in Ref.~\onlinecite{ken2006} an increase of the anisotropy $K$ also increases the temperature range in which the HTI phase is stable, because a larger $K$ competes with the fixed spin length constraint from the quartic terms in the Landau expansion, which becomes stronger with decreasing temperature and induces a transverse spin component ending up in the LTI phase. These theoretical considerations are in perfect agreement with our experimental data: in NVO the HTI phase is stable within a temperature range of 2.8 K, while Co-doping increases the stability range to 3.1 K ($x=0.07$) and 3.2 K ($x=0.1$). \newline
To conclude, light doping of Co$^{2+}$ ions (or Mn$^{2+}$ and Zn$^{2+}$ as reported in Ref.~\onlinecite{muk2010}) influences the delicate balance between the magnetic exchange parameters (due to different spin values of the doped ions as well as modified exchange bond lengths and angles because of different atomic radii) and the single-ion anisotropy (due to different spin-orbit coupling of the doped ions), which is predicted by theory\cite{ken2006} to materialize in the propagation vector of the modulated magnetic structures and the temperature stability range of the HTI phase, which our neutron diffraction and magnetization data nicely show.

\begin{acknowledgments}
This work was supported by the Deutsche Forschungsgemeinschaft
through the Sonderforschungsbereich 608 and partially
supported by the Russian Foundation for Basic Research (project 12-02-01261). The support in the neutron experiment by B. Ouladdiaf is thankfully acknowledged.
\end{acknowledgments}

\bibliography{../../../../Literatur/literatur}

\end{document}